\documentclass[referee,sn-basic]{sn-jnl}

\usepackage{graphicx}%
\usepackage{multirow}%
\usepackage{amsmath,amssymb,amsfonts}%
\usepackage{amsthm}%
\usepackage{mathrsfs}%
\usepackage[title]{appendix}%
\usepackage{xcolor}%
\usepackage{textcomp}%
\usepackage{manyfoot}%
\usepackage{booktabs}%
\usepackage{algorithm}%
\usepackage{algorithmicx}%
\usepackage{algpseudocode}%
\usepackage{listings}%

\theoremstyle{thmstyleone}%
\theoremstyle{thmstyletwo}%

\theoremstyle{thmstylethree}%

\begin{document}


    \title[Hybrid climate modelling]{Embedding machine-learnt sub-grid variability improves climate model biases}



\author*[1]{\fnm{Daniel} \sur{Giles}}\email{d.giles@ucl.ac.uk}

\author[2]{\fnm{James} \sur{Briant}}\email{james.briant.21@ucl.ac.uk}

\author[3, 4]{\fnm{Cyril} \sur{J. Morcrette}}\email{cyril.morcrette@metoffice.gov.uk}

\author[2]{\fnm{Serge} \sur{Guillas}}\email{s.guillas@ucl.ac.uk}

\affil*[1]{\orgdiv{Department of Computer Science}, \orgname{University College London}, \orgaddress{ \city{London}, \country{United Kingdom}}}

\affil[2]{\orgdiv{Department of Statistical Sciences}, \orgname{University College London}, \orgaddress{ \city{London}, \country{United Kingdom}}}

\affil[3]{\orgname{Met Office}, \orgaddress{\city{Exeter}, \country{United Kingdom}}}

\affil[4]{\orgdiv{Department of Mathematics and Statistics and Global Systems Institute}, \orgname{University of Exeter}, \orgaddress{\city{Exeter}, \country{United Kingdom}}}


\abstract{The under-representation of cloud formation is a long-standing bias associated with climate simulations. Parameterisation schemes are required to capture cloud processes within current climate models but have known biases. We overcome these biases by embedding a Multi-Output Gaussian Process (MOGP) trained on high resolution Unified Model simulations to represent the variability of temperature and specific humidity within a climate model. A trained MOGP model is coupled in-situ with a simplified Atmospheric General Circulation Model named SPEEDY. The temperature and specific humidity profiles of SPEEDY are perturbed at fixed intervals according to the variability predicted from the MOGP. Ten-year predictions are generated for both control and ML-hybrid models. The hybrid model reduces the global precipitation bias by 18\% and over the tropics by 22\%. To further understand the drivers of these improvements, physical quantities of interest are explored, such as the distribution of lifted index values and the alteration of the Hadley cell. The control and hybrid set-ups are also run in a plus 4K sea-surface temperature experiment to explore the effects of the approach on patterns relating to cloud cover and precipitation in a warmed climate setting.}




\maketitle

General Circulation Models (GCMs) play a vital role in our understanding of climate dynamics and how it is changing worldwide. The state-of-the-art GCMs are used to investigate the effects of different forcing pathways on the evolving climate and the simulation results therein have formed the key basis of the Intergovernmental Panel on Climate Change (IPPC) reports \cite[]{ipcc_ar6_spm}. Despite the state-of-the-art GCMs being optimised to run on some of the largest high performance computing infrastructures, the need for long integration periods limits the spatial resolution afforded. Climate simulations are typically carried out using spatial resolutions of $\mathbf{O}$(100km), which is coarser than those used for global $\mathbf{O}$(10km) and regional $\mathbf{O}$(1km) weather forecasting. The afforded resolution of GCMs has contributed to some of their long standing biases, in particular biases associated with cloud formation, convection, precipitation and interactions between the water cycle and the large-scale dynamics. In contrast higher-resolution kilometre-scale atmospheric models are better at representing these atmospheric processes \cite[]{krueger,lean_high_res_um,kendon2017,slingo22}. Furthermore, there are now cloud- and storm-resolving GCMs being tested, but their temporal integration window is limited. 

The size of a grid-box within a GCM defines the scales at which meteorological dynamics can be modelled. Indeed, only one value is used as the point estimate for each atmospheric variable in each grid-box. There are many atmospheric processes, such as radiative transfer, precipitation growth, cloud formation and convective development that have an affect on the resolved scale, but that are modulated by interactions occurring on scales smaller than the grid. Parametrization schemes have been developed to represent these processes \cite[e.g.][]{stensrud}. Although many schemes include some representation of sub-grid variability in calculating how a given atmospheric process will modify the thermodynamic state of the column, the inputs to the schemes consist of the vertical profiles of variables representing a horizontal mean over the size of the grid-box \cite[e.g.][]{smith90,pbm03,tompkins2005,tke_rhcrit}. Our approach builds upon these methods by explicitly capturing the sub-grid variability to. 

Improving the known biases in climate model simulations is an area of ongoing research with extensive efforts placed on improving parameterization schemes, introducing stochastic components, pushing the resolution envelop by leveraging the latest in high performance computing and/or by integrating data-driven approaches in a hybrid mode \cite[e.g.][]{Ukkonen2020, Lagerquist2021}. As our work focuses on the hybrid climate modelling set up, relevant work is presented here. \cite{Beucler2023} provides a recent overview of machine learning methods being developed for cloud and climate modelling applications, \cite{bb2018,bb2019} used kilometre-scale aqua-planet simulations, and the spatial averaging of their data to the scale of a climate model grid-box, to infer the temperature and humidity tendencies that would need to be added to the coarse model to make it behave more like the high resolution one. They used machine learning techniques to predict the corrections that would need to be applied and then coupled these to their global model leading to significant improvements in its fidelity. This work was extended by \cite{bretherton22} using more realistic high-resolution simulations, that included land and orography, but for up to 40 days ahead forecasts, not climate simulations. \cite{Shamekh2023} capture sub-grid cell cloud structure and organisation by training a neural network which is informed by an organisation metric to predict the parameterised coarse-grained precipitation. This results in an improved prediction skill of precipitation statistics and extremes in a coarse climate model. \cite{Arcomano2022, Arcomano2023} couple a low resolution GCM with a reservoir computing machine learning model, which improved systematic errors and enabled the low resolution climate models to capture phenomenon previously absent, such as the El Ni\~no-Southern Oscillation.

\begin{figure}[bt]
\centering
\includegraphics[width=\columnwidth]{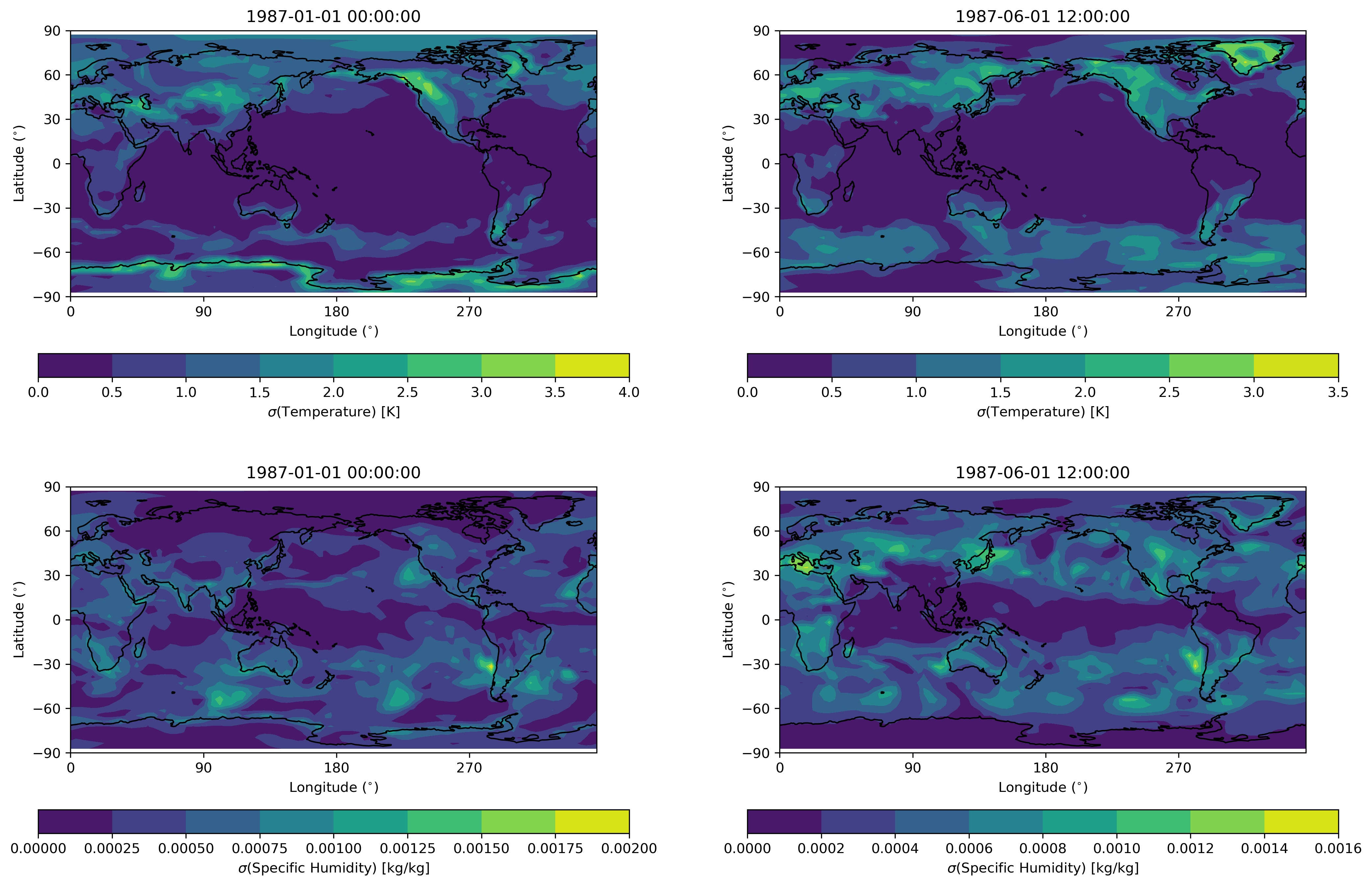}
\caption{Examples of MOGP predictions of temperature and humidity standard deviation on 00:00UTC 1 January and 12:00UTC 1 June 1987 at 925 hPa.
}
\label{fig:mogp-prediction-maps}
\end{figure}

This study explores the use of coarse-grained high-resolution simulations and machine learning to improve the fidelity of a coarse-resolution climate model. Differing from \cite{bb2018,bb2019}, the high-resolution simulation data is not used here to infer and learn a nudging term that is applied to the coarse model. Instead, a novel approach to applying stochastic perturbations is explored. Rather than perturbing the temperature (T) and specific humidity (Q) tendencies coming out of a parametrization scheme \cite[e.g. ][]{buizza1999,christensen2017} or perturbing uncertain parameters within a certain scheme \cite[e.g. ][]{bowler2008,jankov2019}, the high-resolution data is used as training data to quantify the temperature and humidity variability that occurs on scales smaller than the coarse-model grid and the thermodynamic profiles are perturbed within that predicted envelope. 
A Multi-Output Gaussian Process (MOGP) is used to predict the profiles of temperature and humidity variability. 
Key benefits of MOGP models over traditional neural network based approaches are more stability when sampling outside of training distribution, fewer training samples and crucially the inherent uncertainty estimate \cite[]{watson2021machine}. The MOGP model is trained on coarse-grained high-resolution outputs from the UK Met Office's Unified Model \cite[]{ga3p0}. 80 regional patches which spatially cover the globe evenly are simulated for 10 days and with a spatial resolution of 1.5km. Details on the high-resolution model runs and the coarse-graining procedures are outlined in Section \ref{UM_data}. Further, validation results of the trained MOGP are outlined in Section \ref{methods:MOGP}. Figure \ref{fig:mogp-prediction-maps} showcases the patterns of variation that the trained MOGP model predicts in the temperature and specific humidity fields at 925hPa at two different time points which fall within the 10 year experimental window (00:00 UTC 1st January and 12:00 UTC 1st June 1987). For both time points, the MOGP predicts larger variability in temperature over land than the ocean and the effect of the seasons can be seen in the specific humidity variability estimates. Synoptic wave patterns can also be seen for both variables. It should be noted that the MOGP model operates on a GCM column-by-column basis and therefore has not explicitly learned these spatial covariances but can clearly predict spatially coherent patterns that vary in time and space.

In this work the trained MOGP model is two-way coupled with a simplified GCM,  SPEEDY. SPEEDY (Simplified Parameterizations primitivE-Equation DYnamics) is an atmospheric GCM which consists of a spectral primitive-equation dynamic core along with a set of simplified physical parametrization schemes \cite[]{molteni2003}. The prognostic variables are the zonal and meridional components of wind ($u,v$), temperature ($T$), specific humidity ($Q$) and surface pressure ($p_{s}$). A T30 grid resolution is used yielding an approximate grid length of 333km at the equator. The model has 8 vertical layers, defined by sigma levels, where the pressure is normalized by the surface pressure ($p/p_{s}$). SPEEDY is used here to explore the feasibility and impact of two-way coupling of machine-learnt elements into a GCM in order to affect its dynamical evolution and provides motivation for extending this work within a fully-fledged GCM.

Details on the two-way coupling of SPEEDY with the trained MOGP can be found in Section \ref{methods:coupling}. The resultant impact of this fusion of data-science techniques to represent sub-grid thermodynamic variability within a climate model is further explored here. The precipitation outputs from a 10-year hybrid and control simulation are compared against ERA5 \cite[]{era5} to showcase the improvements of the proposed technique (Section \ref{sec:precip}), details on the experimental set-up can be found in Section \ref{methods:experiment}. The physical justifications for the observed improvement in wet and dry biases are explored in Section \ref{sec:physics-drivers}.

Encouraged by the improvements of the hybrid model over the known 10-year run, a warmed climate simulation is also carried out. A 10-year simulation with the sea surface temperature (SST) warmed uniformly by 4$K$ is used as an approximate approach to capturing the dynamics of the warming climate (Section \ref{sec:4ksst}). Increased cloud cover and altered precipitation patterns, which are similar to the findings in Section \ref{sec:precip}, are observed in this warmed 10-year simulation. These results showcase the benefit of the proposed approach for capturing sub-grid scale variability and tackling long-standing GCM biases. Future directions and concluding remarks are outlined in Section \ref{sec:discussion}.

\section{Data generation, coarse-graining and training design}\label{UM_data}
The data used for training are generated using the Met Office Unified Model (MetUM). Global model hindcasts are run from analyses to produce initial and lateral boundary conditions for some limited-area models (LAMs). The global driving model uses the Global Atmosphere 6 configuration \cite[GA6,][]{ga6p0}, which has been used for operational global forecasting. 
 
The LAMs use a kilometre-scale model configuration, which has also been used for operational forecasting \cite[]{bush_ral1}.
The key difference between the LAM and the driving model is that the LAM is run without a convection scheme.
This configuration has been shown to perform better than the coarser model, using a convective parametrization, when simulating
convection over the United Kingdom \cite[]{lean_high_res_um} or over tropical western Africa \cite[]{stein2015}.

This combination of a global and regional versions of the MetUM is known to perform well when located in various regions around the world and 
has been used to study a range of atmospheric processes on the kilometer-scale, including severe weather over New Zealand \cite[]{webster_nesting_suite}, 
strong winds over Antarctica \cite[]{orr2013}, 
convective systems over the American Midwest \cite[]{hwt2017} and South Africa \cite[]{keat2019}, and precipitation over South Korea \cite[]{song2018}. 

\begin{figure}[bt]
\centering
\includegraphics[width=\columnwidth]
{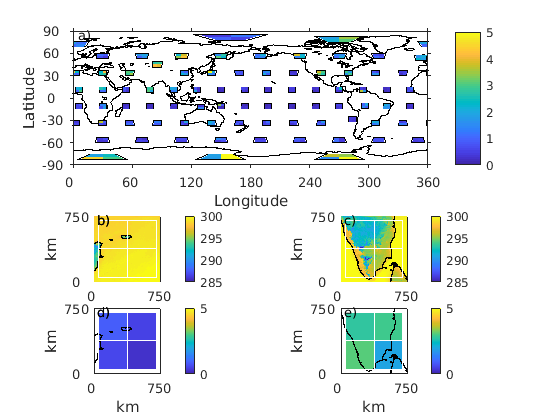}
\caption{a) Location of the 80 limited-area models (LAMs) across the globe.
Each LAM is then split into a 2x2 array, where coarse-graining is performed. An average or a standard deviation can be calculated over these patches, as shown by the shading which, as an example, represents the standard deviation of near-surface temperature [K] at 00Z on 1 Jan 2020. Examples of the 1.5 km surface temperature data are shown for b) the Arabian Sea and c) Southern India and Sri Lanka, while examples of their standard deviation are shown in d) and e).}
\label{fig:quad_icosahedral_map}
\end{figure}

In this study, the nesting suite is extended to cope with 80 separate LAMs, each defined to be 512 $\times$ 512 grid-points in the horizontal with a grid-length of 1.5 km.
The locations of the 80 nested LAMs are shown by the black polygons in Fig. \ref{fig:quad_icosahedral_map} and these were chosen to ensure an even distribution over the Earth's sphere. For consistency, all the LAMs use the same Regional Atmosphere and Land configuration \cite[RAL1-T,][]{bush_ral1} \footnote{The use of the RAL1-M configuration is avoided as it applies stochastic temperature perturbations near the surface. Although this improves simulations of convective showers, it  leaves an imprint on the surface temperature variability.}. 
The global driving model is re-initialised every 24 hours using 00Z operational MetUM atmospheric analyses and each of the 80 nests is free-running, with 3-hourly updated lateral boundary conditions, for 10 days covering the start of January 2020. Global daily sea surface temperatures are provided by the OSTIA system \cite[]{ostia}. The LAM diagnostics are outputted 4 times per day (00:00, 06:00, 12:00 and 18:00 UTC).

Within each $512 \times 512$ LAM, the outer 32 grid-points are treated as a spin-up region and ignored. The remaining $448 \times 448$ region is cut up into 4 sub-regions each of $224 \times 224$ grid-points. With the LAM grid-size of 1.5 km, these are 336 km wide. The SPEEDY model has a latitude-longitude grid consisting of 48 $\times$ 96 points so that the square root of the mean grid-box area is 333 km. Therefore the grid-box sizes in SPEEDY are very similar in size to the LAM sub-regions.

Temperature, specific humidity and pressure data from each sub-region, in each LAM, are processed to produce profiles. This processing consists of calculating horizontal means and standard deviation at each height in the vertical. The sub-grid variability in temperature and humidity in the training data varies due to the local meteorology from place to place. As an example, the standard deviation of temperature for each sub-region at the model level nearest to the surface is shown by the shading in Fig. \ref{fig:quad_icosahedral_map} (a).

Figure \ref{fig:quad_icosahedral_map} (b-e) provide examples of the high-resolution simulations that form the training data by focussing on the temperature at the model level nearest to the surface in two contrasting but nearby locations. Panel (b and d) are for the LAM in the Arabian Sea (located off the north-east coast of Somalia) while
panels (c and e) are for the LAM covering southern India and parts of Sri Lanka. 
The colour shading in panels (b) and (c) shows the original high-resolution fields, and the squares (located 32 grid-boxes in from the lateral edges) indicate the regions over which the horizontal mean and standard deviations are calculated. Panels (d) and (e) showcase the standard deviation of the temperature calculated over the sub-grid boxes.

The training data used for the MOGP are sampled from a design which enforces coverage of each sub-region of the 80 LAMs. The 320 regions (80 LAMs with each having four sub-regions) are each sampled twice randomly in time to form a training dataset size of 640 vertical profiles. The vertical profiles are sampled at sigma coordinates which correspond to the 8 levels defined in SPEEDY. Full details on the training and validation of the MOGP model can be found in Section \ref{methods:MOGP}.

\section{Global mean precipitation over 10-year simulation (1982-1992)} \label{sec:precip}

\begin{figure}[bt]
\centering
\includegraphics[width=\columnwidth]{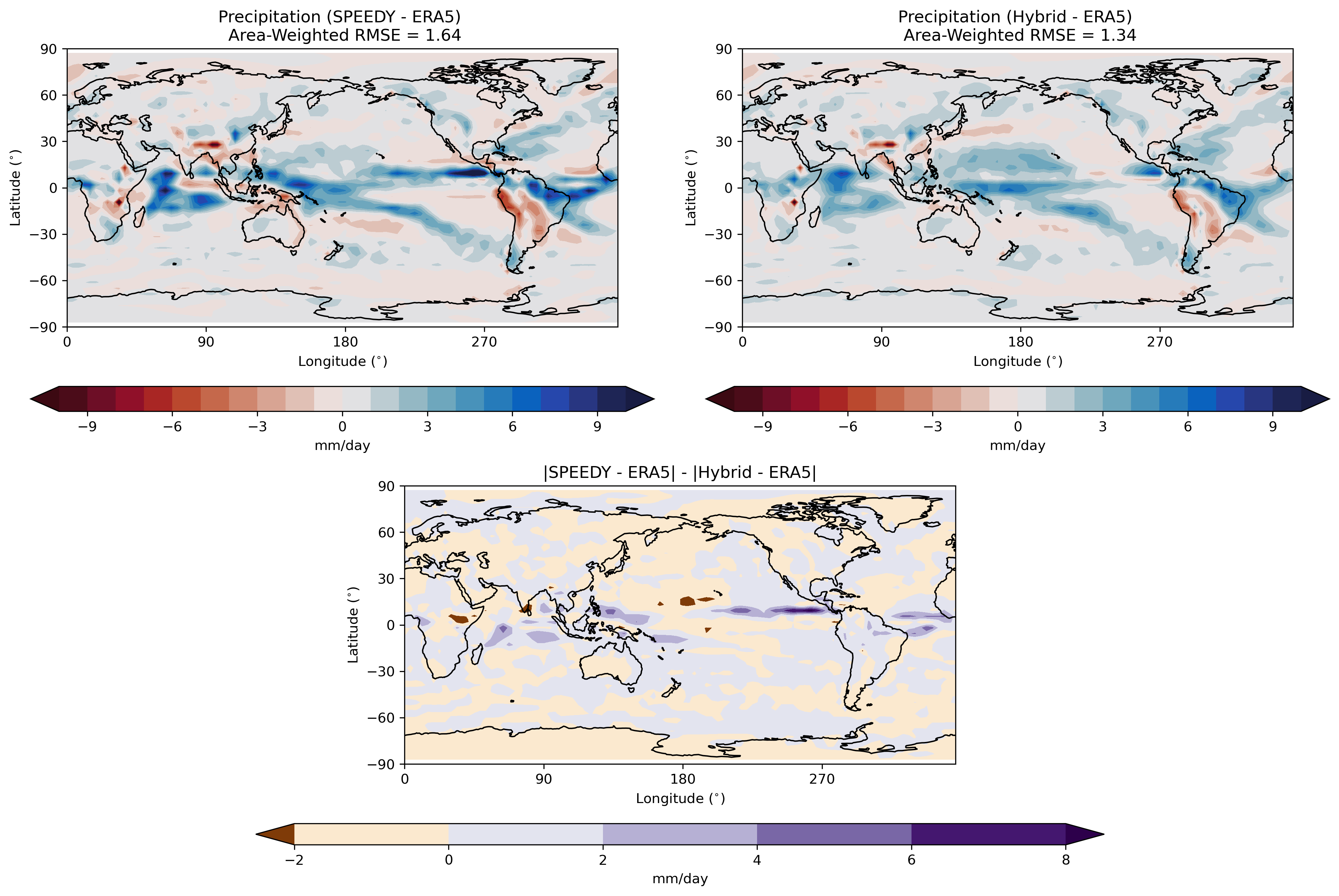}
\caption{Maps of precipitation error between the control (top left) and the hybrid experiment (top right) versus the ERA5 data for the same period. The weighted RMSE values are included in the titles. Bottom centre: the difference of the absolute errors of the control and hybrid runs.
}
\label{fig:era5-precip}
\end{figure}

A 10-year simulation of the global atmosphere is carried out using both the control (SPEEDY) and hybrid (SPEEDY and MOGP) models. The 10-year period covers 1st January 1982 to 1st January 1992. The resultant mean precipitation is then compared to the European Centre for Medium-Range Weather Forecasts (ECMWF) Atmospheric Reanalysis (ERA5) dataset for the same period. To ensure consistency in the results three independent hybrid simulations are carried out, details on the experimental set-up are outlined in section \ref{methods:experiment}. As seen in Figure \ref{fig:era5-precip}, the hybrid approach improves upon the overall bias of the control simulation run. Figure \ref{fig:era5-precip} showcases the mean precipitation from one of the hybrid runs (randomly selected). A mean 18\% reduction in the global area-weighted RMSE and a mean 22\% reduction in the area-weighted RMSE within the tropics region ($\pm 23.5 ^{\circ}$ latitude) is calculated across three hybrid runs (Table \ref{tab:RMSE}). To test the utility of the MOGP predictions, naively perturbed runs form a baseline model. Stochastic noise is introduced at the same time intervals as the hybrid model and various levels of noise are trialled. Full details on the implementation of the naive stochastic models can be found in Section \ref{method:naive}. The mean across three naively perturbed runs with $1\%$ noise level give a mean global and tropics region area-weighted RMSE improvement of 8.7\% and 10.4\% respectively (Table \ref{tab:RMSE}). Note that selecting $5\%$ noise level for the naively perturbed runs results in a degradation in the performance. To give a sense of the magnitude of these bias reductions, the massive update from CESM1 to CESM2 \cite[]{danabasoglu2020community} resulted in a reduction of the RMSE in precipitation of around 20\%, with all atmospheric parameterizations updated across several teams over October 2015 to April 2018.  Areas of particular change in Figure \ref{fig:era5-precip} are the reduction in error in the Arabian Sea, central South America and off the west coast of Central America. Figure \ref{fig:era5-precip} (bottom) showcases the difference in the absolute errors between the hybrid and control run, note the difference in positive and negative range. As a whole, the hybrid model improves upon the biases of the control run.

\begin{figure}[bt]
\centering
\includegraphics[width=\columnwidth]{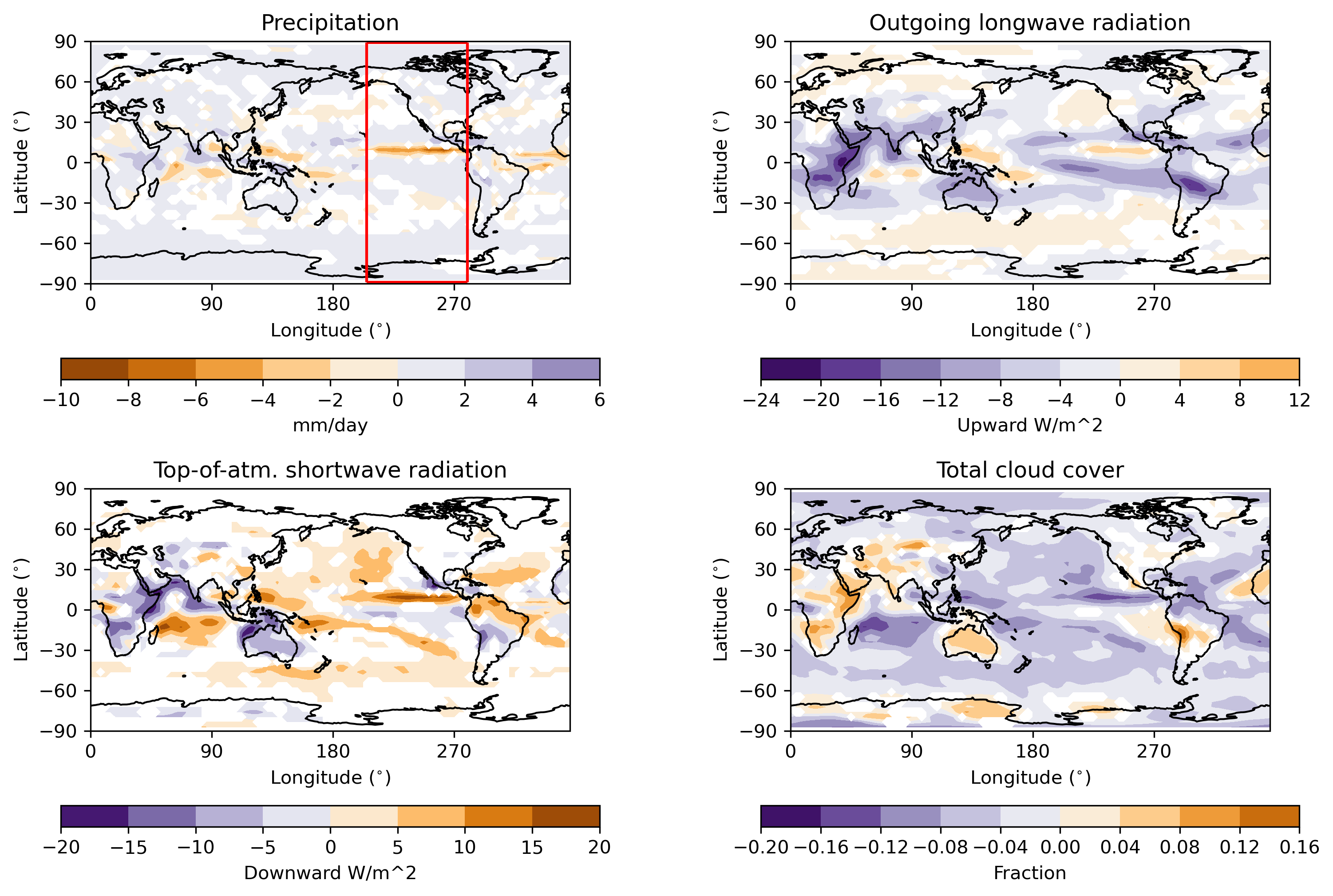}
\caption{Differences in mean precipitation, top-of-atmosphere outgoing long-wave radiation, top-of-atmosphere short-wave radiation and cloud cover between the hybrid and control SPEEDY simulations. Statistically insignificant results are masked.}
\label{fig:field-diffs}
\end{figure}

To further explore these results Figure \ref{fig:field-diffs} highlights the differences between mean precipitation, outgoing long-wave radiation, top-of-the-atmosphere downward solar radiation and cloud cover between the hybrid and control run. In all four subplots a pixel-wise Welch's t-test (2$\sigma$ level) has been used to mask regions (shaded white) that exhibit statistically insignificant changes. This operation ensures the subsequent analyses consider only the true signal of the changes made by our hybrid method. Welch's method is chosen over Student's t-test as Welch's allows for unequal variance between the distributions. 

The areas over land with increased precipitation correspond to an increase in cloud cover, a reduction in top-of-the-atmosphere downward solar radiation (more reflection) and less outgoing long-wave radiation. This points towards more optically thick and deep clouds being produced in the hybrid set-up.

\section{Physical explanation for precipitation differences}\label{sec:physics-drivers}
The perturbations that the hybrid model introduce are informed by the high-resolution variability but are on average zero, so further investigation into the physical drivers of the differences are explored.

\subsection{Hadley cell circulation}
\begin{figure}[bt]
\centering
\includegraphics[width=0.3\columnwidth, height = 0.3\columnwidth]{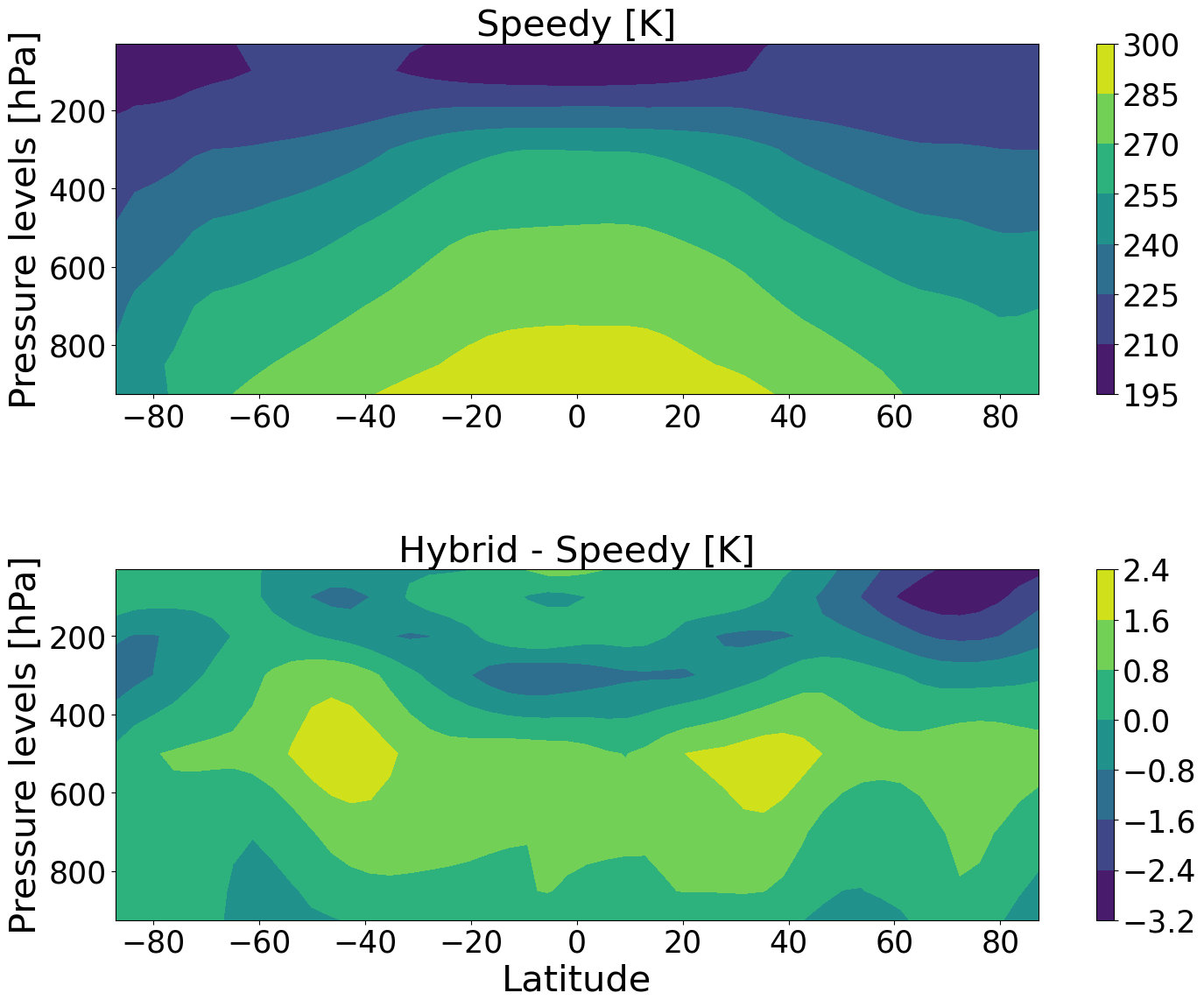}
\includegraphics[width=0.45\columnwidth]{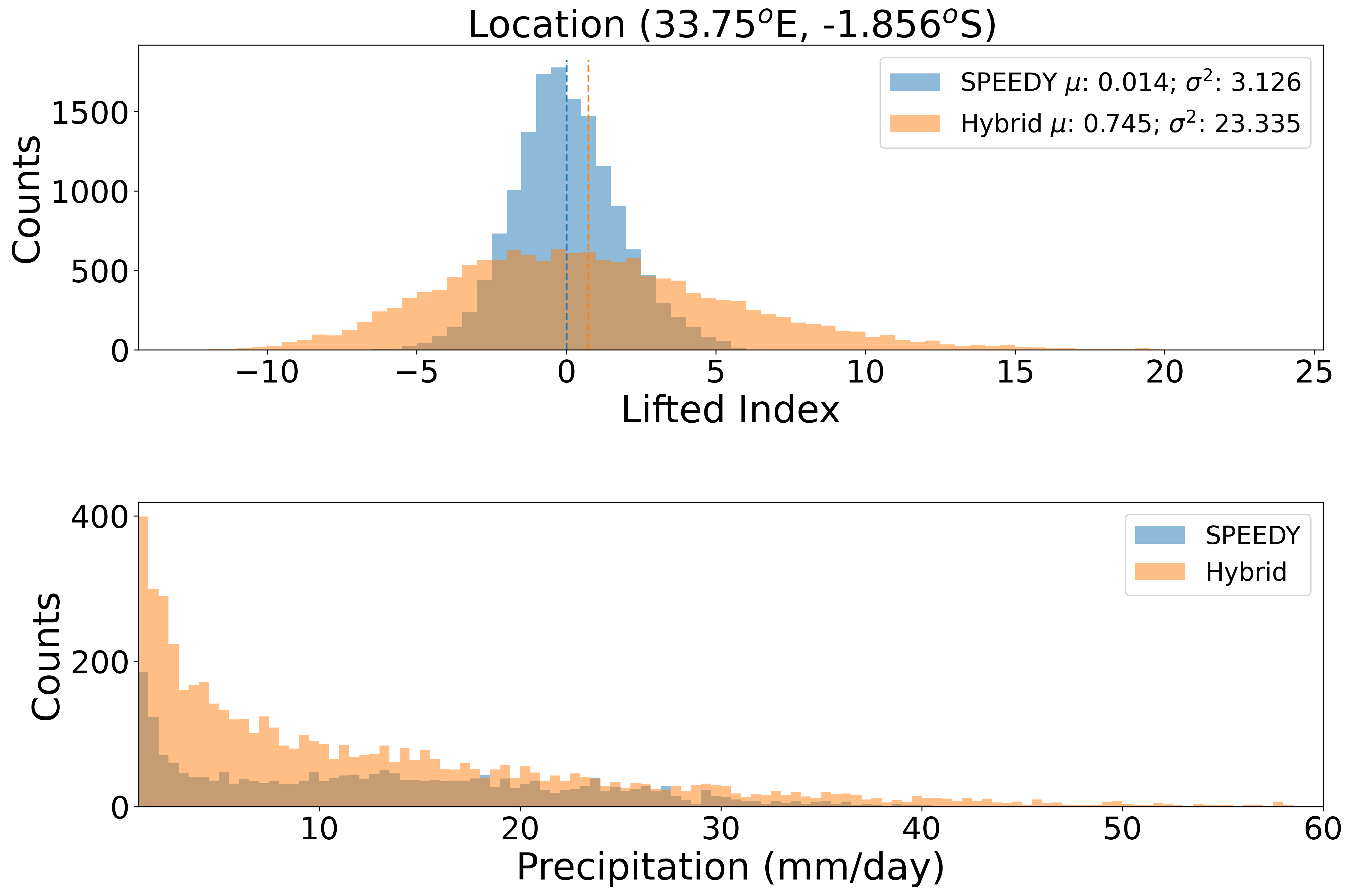}
\caption{Top Left: Zonal averages of the temperature over the limited area (168.75$^{\circ}$ W to 78.75$^{\circ}$ W ) in the Pacific for the control run. Bottom Left: Difference of the zonal averages of the temperature between the control and hybrid run, see top left subplot of Fig. \ref{fig:field-diffs} for the bounding box. Top Right: Lifted index histograms for both the control and hybrid simulation in central Africa (33.75$^{\circ}$ E, -1.856 $^{\circ}$ S). Bottom Right: Precipitation histogram with the zero counts removed.
}
\label{fig:physics}
\end{figure}

The Hadley cell is a major atmospheric circulation pattern that dominates the precipitation and cloud coverage patterns of the tropics and sub-tropics \cite[e.g. ]{Vallis_2017}. It is a system of air circulation, driven by the temperature difference between the equatorial and subtropical regions. It plays a vital role in global precipitation patterns and the position of its Intertropical Convergence Zone (ITCZ) migrates pole-wards with the seasons. Figure \ref{fig:physics} (left) plots the difference between the zonal averages of temperature over the limited area of the Pacific region, see Figure \ref{fig:field-diffs} (top left) for the red bounding box. The hybrid run exhibits a weakened temperature gradient between the equator and the poles when compared to the control run. This weakened temperature gradient contributes to a weakened Hadley circulation and therefore a reduction in the precipitation off the west coast of central America, as seen in Figure \ref{fig:field-diffs} (top left).

\subsection{Lifted index and extreme events}
The lifted index (LI, \cite{galway1956}) quantifies the temperature difference between a rising surface-based parcel and the environment in the mid-troposphere. As such it is a measure of the potential vigour of convection. The lifted-index is calculated every 6 hours for both the hybrid and control run for a point in central Africa. This point is chosen as it exhibits an increase in mean precipitation for the hybrid run. The histograms of the lifted index for this point are plotted in the right top subplot in Figure \ref{fig:physics}. There are clear differences in the histograms obtained. The presence of a larger number of negative LI values, where LI $\leq$ -4, points towards a larger number of unstable atmospheric conditions. The unstable atmospheric conditions result in a larger number of extreme precipitation events, as can be seen in the bottom right subplot of Figure \ref{fig:physics}. These histograms reveal the effect that the perturbations have on the nonlinear characteristics of the parameterization schemes. Despite the perturbations being mean zero the nonlinear nature of convection results in a large change in the time-averaged thermodynamics and in the precipitation coming from these systems.

\section{Plus 4K sea-surface temperature simulation}\label{sec:4ksst}
\begin{figure}[bt]
\centering
\includegraphics[width=\columnwidth]{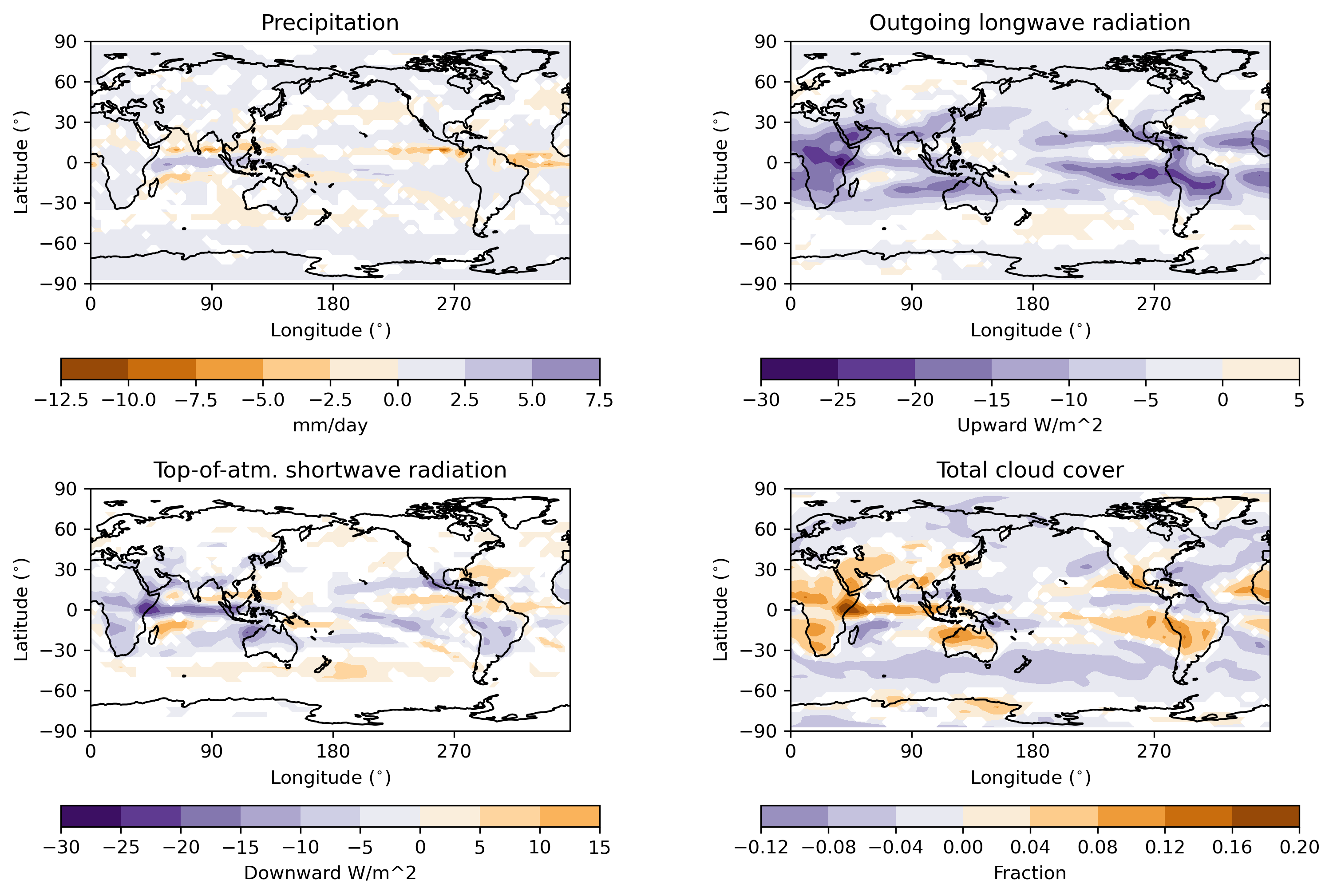}
\caption{Differences in precipitation, outgoing long-wave radiation, top-of-atmosphere solar radiation and cloud cover between the hybrid and control SPEEDY simulations in the plus 4K sea surface temperature simulation.}
\label{fig:warm_4K}
\end{figure}

Motivated by the changes and bias reductions that the hybrid approach introduces over the historical period a warmed climate simulation is carried out to explore the effects of using the same trained MOGP model on a warmed scenario. SPEEDY does not allow for the easy integration of different emission pathways and ocean coupling therefore a 10-year simulation with a plus 4 K sea-surface temperature (SST) increase is carried out. \cite{Qin2022} provides justification for this choice as they compared coupled to plus 4K SST experiments across the CMIP5 and CMIP6 models and concluded that atmosphere-only experiments are sufficient when investigating climate feedbacks. Similar to Figure \ref{fig:field-diffs}, Figure \ref{fig:warm_4K} showcases the differences between 10-year mean precipitation, outgoing long-wave radiation, top-of-the-atmosphere downward solar radiation and cloud cover fields for the hybrid and control plus 4K SST simulations. It can be seen that compared to the no climate change scenario in Figure \ref{fig:field-diffs} there is a stronger signal of reduced precipitation over Pakistan/Northern India and the US Pacific West Coast. The cloud cover shows very large differences over Africa, demonstrating the capacity of our method to better enable the manifestation of a climate change signal. Similar to Figure \ref{fig:field-diffs} there exists a notable reduction in the outgoing long-wave radiation in the hybrid run which again points towards more optically deep clouds being produced by the hybrid approach. This could have large implications for a fully fledged climate change simulation where the insulating effect of the optically deep clouds can influences climate feedbacks.

\section{Discussion}\label{sec:discussion}

SPEEDY is a simplified model. As such, it would be relatively easy to improve its fidelity by increasing its horizontal and vertical resolution or by using a more sophisticated set of parametrization schemes. Although SPEEDY is a simplified model, it is interesting to note that in terms of precipitation, it has a wet bias over the Arabian Sea and a dry bias over the Bay of Bengal. This bias is both very common in GCMs and long-standing. It is seen in the multi-model mean over several iterations of climate model inter-comparison studies \cite[]{tian20}. It is also a pattern of precipitation bias which has been present in several versions of the Met Office global climate model \cite[]{ga7p0}.

The goal here is to use SPEEDY, which is much cheaper to run than state-of-the-art GCMs, but which exhibits some biases common to those more expensive GCMs, as a testbed for how the introduction of machine-learnt thermodynamic variability can impact on a model's climate. The 18\% reduction in area-weighted RMSE we observed with the hybrid SPEEDY is an illustration of the potential that results from our approach, here for SPEEDY and later for other models.

In these simulations, the perturbations have zero mean  
but have a non-zero impact on the model climate. Specifically, in situations of tropical weather, if the profile is neutral, and hence on the cusp of stable or unstable, then a small perturbation can push it to being stable, leading to no rain, while a similar magnitude perturbation in the opposite direction can lead to an unstable situation which would lead to a lot of precipitation. It is shown here that despite being zero on average, the perturbations do change the patterns of precipitation. These changes occur both through the immediate, local, changes due to whether convection occurs, but also as a result of the feedback that this change in convective activity has on the synoptic scale and hence the large-scale flow. 

Several avenues present themselves for future work. 
Firstly, the MOGP could be retrained using a larger dataset, which has been extended to consist of more high-resolution simulations from more locations and for longer periods. This could include data from kilometre-scale global simulations \cite[e.g. ][]{slingo22}.

Secondly, the MOGP could be coupled to more complex GCMs to confirm that the approach described here leads to similar changes in the global distributions of simulated precipitation.

Finally, the manner in which the predicted uncertainty is sampled should be investigated. At present, the mean profiles of T and Q are perturbed based on the predicted standard deviation, with a correlation imposed on the samples drawn at each height. Although some coherence between T and Q perturbations at each height is possible and some vertical coherence is also likely, it is also reasonable to assume that T and Q perturbations may not always be perfectly correlated or that some vertical decorrelation should be allowed. As a result, it would be interesting to study how these correlations affect the results. These correlations are likely to depend on the details of the local meteorology. Most interestingly, it would be useful to investigate how to make these correlations be something that is learnt from the high-resolution training data.

\backmatter

\bmhead{Acknowledgments}

SG and DG were funded by EPSRC project EP/W007711/1 ``Software Environment for Actionable \& VVUQ-evaluated Exascale Applications" (SEAVEA). JB was funded by EPSRC project EP/V520263/1 ``Maths DTP 2020 University College London".

\section{Online Methods}
\label{sec:method}

\subsection{MOGP}\label{methods:MOGP}
Profiles of the variances of the temperature and specific humidity are predicted using a Multi-Output Gaussian Process \cite[MOGP, ][]{mogp}. 
Examples of the successful implementations of MOGPs and discussions of their fitting and predictions are given by \cite{gopinathan2021probabilistic}, 
\cite {salmanidou2021probabilistic} 
and \cite{giles2021faster}.

The MOGP predicts the variances on a column by column basis and takes as input 
the land-sea fraction, mean orography, standard deviation of the orography, mean surface-level pressure and mean air temperature
and mean specific humidity at each of the 8 height levels used in SPEEDY. 

\subsubsection{Training}
\label{sec:mogp-training}
The training dataset consists of 12,800 profiles: $80 \times 4 = 320$ sub-regions, 4 times each day (00, 06, 12 and 18 UTC) for 10 days. The training profiles are chosen randomly over the time dimension but with exactly 2 training points for each sub-region. This approach ensures an even coverage globally of all the available orographies in the training dataset. Sophisticated experimental design algorithms and sampling methods cannot be easily implemented here as the sampling space is discrete and the notion of a space-filling design is not obvious. 

The training is carried out using a workstation ( 16 x Intel(R) Core(TM) i7-9800X CPU), with the hyperparameters estimated using a LFBGS solver. A Mat\'ern-5/2 kernel function is chosen to train the MOGP due to its versatility. As there is a large difference in the scales of the predicted outputs, the standard deviation on temperature is order of magnitudes larger than the standard deviation on the specific humidity, the specific humidity standard deviation is scaled by 1000.

\subsubsection{Validation}
\label{sec:mogp-validation}
\begin{figure}[bt]
\centering
\includegraphics[width=0.9\columnwidth, height=0.7\columnwidth]{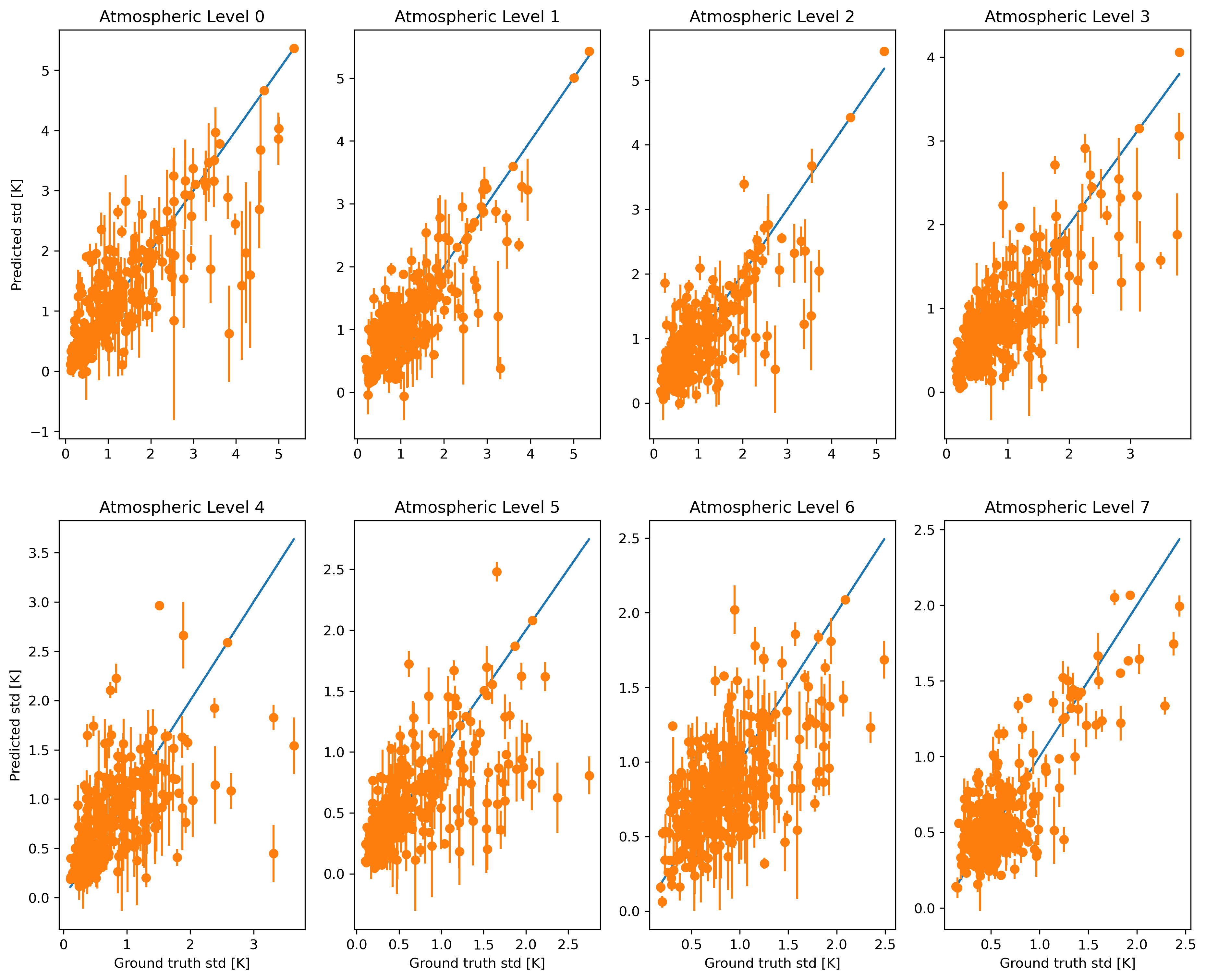}
\caption{MOGP validation on test dataset. The predicted temperature standard deviation is plotted against the ground truth value for each level of the SPEEDY atmosphere.}
\label{fig:mogp-predictions-T}
\end{figure}

\begin{figure}[bt]
\centering
\vspace{1mm}
\includegraphics[width=0.8\columnwidth, height=0.8\columnwidth]{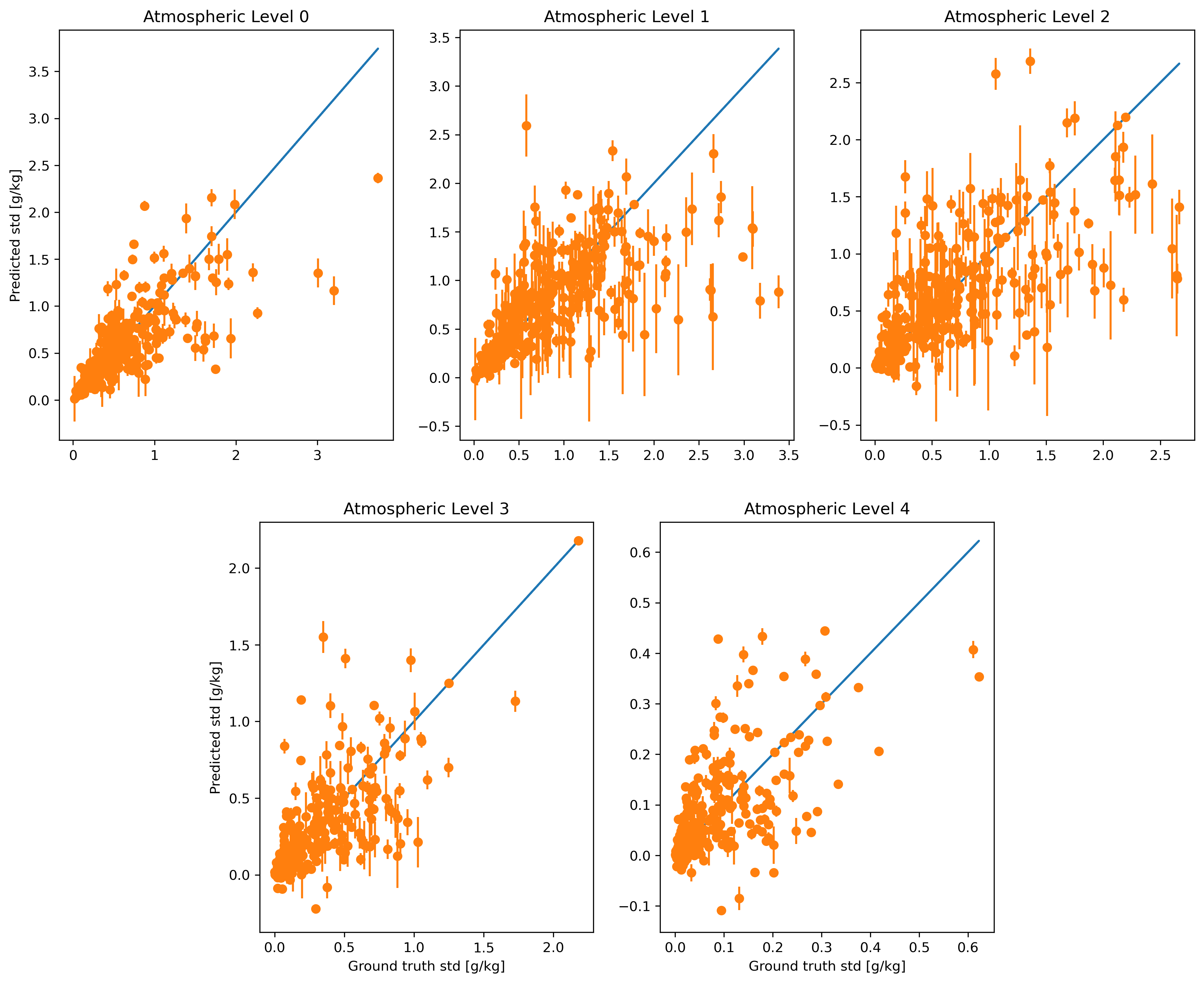}
\caption{MOGP validation on test dataset. The predicted specific humidity standard deviation is plotted against the ground truth value for each level of the SPEEDY atmosphere.}
\label{fig:mogp-predictions-Q}
\end{figure}

To ensure the MOGP has been trained successfully, its predictions are compared to ground truth values on a test dataset. In Figs. \ref{fig:mogp-predictions-T} and \ref{fig:mogp-predictions-Q} the predicted versus ground truth values can be seen. The uncertainty on the MOGP predictions are included as error bars. It can be seen that the MOGP captures the overall trend but is over confident in its predictions, especially for the specific humidity estimates.


\subsection{Two-way coupling of GCM and MOGP}\label{methods:coupling}
The two-way coupling of SPEEDY and the trained MOGP relies on I/O and is implemented as follows.
After each 6 hour forward integration of SPEEDY all the column profiles of T and Q are written to disk. These grid-box mean profiles, the land-sea mask, the grid-box mean surface pressure and the grid-box mean and standard deviation of orography form the inputs to the MOGP. The MOGP then predicts profiles of the standard deviations of T and Q. The mean profiles of T and Q are perturbed based on the predicted standard deviation, with a correlation imposed on the samples drawn at each height. The new profiles of T and Q are then written back to disk. The perturbed profiles are read in by SPEEDY, where another forward integration occurs and the cycle repeats. 
The coupling of the two models is managed by a python wrapper, the code for which is available \footnote{A link to our code will be made available upon acceptance}. 

Using the model-predicted profiles of T and Q as inputs, the MOGP is able to produce a range of different predictions of T and Q variability. Figure \ref{fig:mogp-prediction-maps} shows maps of this variability at 925 hPa, as an exmaple. There is generally more sub-grid T variability predicted over land than over sea and in the summer hemisphere than in the winter hemisphere. This makes sense as increased insolation and variable land surfaces and orography will naturally lead to more variability in lower-troposheric temperature in the summer and over land. However, there are also regions of increased T variability over the ocean, especially in the storm track regions. Humidity variability is larger in the tropics, where humidity is generally larger, but this is not uniform.
Synoptic variability is also apparent as waves patterns in the both T and Q variability. 
Fig. \ref{fig:mogp-prediction-maps}  illustrates that the MOGP is able to produce a range of predictions and that it generates T and Q variability that is itself variable in space and time.

\subsection{Experimental set up}\label{methods:experiment}
A 1-year control simulation starting from $1^{st}$ January 1981 provides spin-up initial conditions for the control and hybrid runs. The control and hybrid runs are then integrated forward for a further 10 years ($1^{st}$ January 1982 to $1^{st}$ January 1992). The hybrid experiment consists of SPEEDY two-way coupled to the MOGP. SPEEDY's predicted T and Q profiles are perturbed within their predicted standard deviation at 6 hourly intervals and various diagnostic fields are outputted. The hybrid simulations are repeated three times to ensure that the results are consistent. The area weighted root mean squared error (RMSE) are calculated against ERA5 precipitation data covering the same period. The breakdown of the RMSE values are displayed in Table \ref{tab:RMSE}. 

\subsection{Naive stochastic perturbations}\label{method:naive}
Perturbing the profiles in a naive manner acts as the baseline model to compare against. In this naive approach the profiles are perturbed by using the same coupling framework as in section \ref{methods:coupling}. However, instead of predicting the standard deviation of T and Q at each level using the MOGP model the profiles are perturbed by additive Gaussian noise given by $\mathcal{N}(0,\,\epsilon \mu_{T/Q})$, where $\mu_{T/Q}$ is the mean values and  $\epsilon \in \{0.01, 0.05, 0.1\}$. The stochastically perturbed simulation with $\epsilon = 0.1$ fails to run due to non-physical profiles being produced. The area weighted RMSE values for the other runs are given in Table \ref{tab:RMSE}.

\begin{table}[]
\begin{tabular}{|l|p{20mm}|p{20mm}|p{20mm}|p{20mm}|}
\hline
& Global Weighted RMSE & Global Relative Error Difference \% & Tropics Weighted RMSE & Tropics Relative Error Difference \%\\ \hline
Control SPEEDY & 1.6389  &   N/A   & 2.9707 & N/A   \\ \hline
Hybrid Run   & 1.3419  & 18.12 & 2.3341 & 21.43   \\ \hline
Hybrid Run   & 1.3346  & 18.57 & 2.3075 & 22.32  \\ \hline
Hybrid Run   & 1.3347  & 18.56 & 2.3194 & 21.93   \\ \hline
Naive Perturbation ($\epsilon = 0.01$) & 1.4976 & 8.62 & 2.6636 & 10.34 \\ \hline
Naive Perturbation ($\epsilon = 0.01$) & 1.5007 & 8.43 & 2.67 & 10.12\\ \hline
Naive Perturbation ($\epsilon = 0.01$) & 1.4921 & 8.96 & 2.6553 & 10.62\\ \hline
Naive Perturbation ($\epsilon = 0.05$) & 4.8993 & N/A & 8.6131 & N/A \\ \hline
Naive Perturbation ($\epsilon = 0.05$) & 4.8943 & N/A & 8.5979 & N/A \\ \hline
Naive Perturbation ($\epsilon = 0.05$) & 4.8870 & N/A & 8.5806 & N/A\\ \hline
Naive Perturbation ($\epsilon = 0.1$) & N/A & N/A & N/A & N/A \\
\hline
\end{tabular}
\caption{Global and tropics area weighted precipitation RMSE values against ERA5 data for the control SPEEDY, hybrid and naively perturbed runs. For cases of improvement the percentage relative error against the control run are also calculated.}
\label{tab:RMSE} 
\end{table}










\bibliography{refs}

\begin{thebibliography}{46}
\providecommand{\natexlab}[1]{#1}
\providecommand{\url}[1]{{#1}}
\providecommand{\urlprefix}{URL }
\providecommand{\doi}[1]{\url{https://doi.org/#1}}
\providecommand{\eprint}[2][]{\url{#2}}
 \bibcommenthead

\bibitem[{Arcomano et~al(2022)Arcomano, Szunyogh, Wikner, Pathak, Hunt, and
  Ott}]{Arcomano2022}
Arcomano T, Szunyogh I, Wikner A, et~al (2022) {A Hybrid Approach to
  Atmospheric Modeling That Combines Machine Learning With a Physics-Based
  Numerical Model}. Journal of Advances in Modeling Earth Systems 14(3):1--21.
  \doi{10.1029/2021MS002712}

\bibitem[{Arcomano et~al(2023)Arcomano, Szunyogh, Wikner, Hunt, and
  Ott}]{Arcomano2023}
Arcomano T, Szunyogh I, Wikner A, et~al (2023) {A Hybrid Atmospheric Model
  Incorporating Machine Learning Can Capture Dynamical Processes Not Captured
  by Its Physics-Based Component}. Geophysical Research Letters 50(8):1--10.
  \doi{10.1029/2022GL102649}

\bibitem[{Beucler et~al(2023)Beucler, Ebert-Uphoff, Rasp, Pritchard, and
  Gentine}]{Beucler2023}
Beucler T, Ebert-Uphoff I, Rasp S, et~al (2023) Machine Learning for Clouds and
  Climate, American Geophysical Union (AGU), chap~16, pp 325--345.
  \doi{10.1002/9781119700357.ch16}

\bibitem[{Bowler et~al(2008)Bowler, Arribas, Mylne, Robertson, and
  Beare}]{bowler2008}
Bowler NE, Arribas A, Mylne KR, et~al (2008) The mogreps short-range ensemble
  prediction system. Quarterly Journal of the Royal Meteorological Society
  134(632):703--722. \doi{10.1002/qj.234}

\bibitem[{Brenowitz and Bretherton(2018)}]{bb2018}
Brenowitz ND, Bretherton CS (2018) Prognostic validation of a neural network
  unified physics parameterization. Geophysical Research Letters
  45(12):6289--6298. \doi{10.1029/2018GL078510}

\bibitem[{Brenowitz and Bretherton(2019)}]{bb2019}
Brenowitz ND, Bretherton CS (2019) Spatially extended tests of a neural network
  parametrization trained by coarse‐graining. J Adv Model Earth Syst
  11(8):2728--2744. \doi{10.1029/2019MS001711}

\bibitem[{Bretherton et~al(2022)Bretherton, Henn, Kwa, Brenowitz, Watt-Meyer,
  McGibbon, Perkins, Clark, and Harris}]{bretherton22}
Bretherton CS, Henn B, Kwa A, et~al (2022) Correcting coarse-grid weather and
  climate models by machine learning from global storm-resolving simulations. J
  Adv Model Earth Syst 14:e2021MS002794. \doi{10.1029/2021MS002794}

\bibitem[{Buizza et~al(1999)Buizza, Miller, and Palmer}]{buizza1999}
Buizza R, Miller M, Palmer TN (1999) Stochastic representation of model
  uncertainties in the ecmwf ensemble prediction system. Quarterly Journal of
  the Royal Meteorological Society 125(560):2887--2908.
  \doi{10.1002/qj.49712556006}

\bibitem[{Bush et~al(2019)Bush, Allen, Bain, Boutle, Edwards, Finnenkoetter,
  Franklin, Hanley, Lean, Lock, Manners, Mittermaier, Morcrette, North, Petch,
  Short, Vosper, Walters, Webster, Weeks, Wilkinson, Wood, and
  Zerroukat}]{bush_ral1}
Bush M, Allen T, Bain C, et~al (2019) The first {M}et {O}ffice {U}nified
  {M}odel/{JULES} {R}egional {A}tmosphere and {L}and configuration, {RAL1}.
  Geosci Model Dev \doi{10.5194/gmd-2019-130}

\bibitem[{Christensen et~al(2017)Christensen, Lock, Moroz, and
  Palmer}]{christensen2017}
Christensen HM, Lock SJ, Moroz IM, et~al (2017) {Introducing independent
  patterns into the Stochastically Perturbed Parametrization Tendencies (SPPT)
  scheme}. Quarterly Journal of the Royal Meteorological Society
  143(706):2168--2181. \doi{10.1002/qj.3075}

\bibitem[{Danabasoglu et~al(2020)Danabasoglu, Lamarque, Bacmeister, Bailey,
  DuVivier, Edwards, Emmons, Fasullo, Garcia, Gettelman
  et~al}]{danabasoglu2020community}
Danabasoglu G, Lamarque JF, Bacmeister J, et~al (2020) {The Community Earth
  System Model Version 2 (CESM2)}. Journal of Advances in Modeling Earth
  Systems 12(2):e2019MS001916. \doi{10.1029/2019MS001916}

\bibitem[{Daub et~al(2022)Daub, Strickson, and Barlow}]{mogp}
Daub E, Strickson O, Barlow N (2022) Multi-output gaussian process emulator.
  \urlprefix\url{https://github.com/alan-turing-institute/mogp-emulator}

\bibitem[{Donlon et~al(2012)Donlon, Martin, Stark, Roberts-Jones, Fiedler, and
  Wimmer}]{ostia}
Donlon CJ, Martin M, Stark JD, et~al (2012) The operational sea surface
  temperature and sea ice analysis ({OSTIA}) system. Remote Sensing of
  Environment 116:140--158. \doi{10.1016/j.rse.2010.10.017}

\bibitem[{Galway(1956)}]{galway1956}
Galway JG (1956) The lifted index as a predictor of latent instability. Bull
  Amer Meteor Soc 37:528--529. \doi{10.1175/1520-0477-37.10.528}

\bibitem[{Giles et~al(2021)Giles, Gopinathan, Guillas, and
  Dias}]{giles2021faster}
Giles D, Gopinathan D, Guillas S, et~al (2021) Faster than real time tsunami
  warning with associated hazard uncertainties. Frontiers in Earth Science
  8:597865. \doi{10.3389/feart.2020.597865}

\bibitem[{Gopinathan et~al(2021)Gopinathan, Heidarzadeh, and
  Guillas}]{gopinathan2021probabilistic}
Gopinathan D, Heidarzadeh M, Guillas S (2021) Probabilistic quantification of
  tsunami current hazard using statistical emulation. Proceedings of the Royal
  Society A 477(2250):20210180. \doi{10.1098/rspa.2021.0180}

\bibitem[{Hersbach et~al(2020)Hersbach, Bell, Berrisford, Hirahara, Horányi,
  Muñoz-Sabater, Nicolas, Peubey, Radu, Schepers, Simmons, Soci, Abdalla,
  Abellan, Balsamo, Bechtold, Biavati, Bidlot, Bonavita, De~Chiara, Dahlgren,
  Dee, Diamantakis, Dragani, Flemming, Forbes, Fuentes, Geer, Haimberger,
  Healy, Hogan, Hólm, Janisková, Keeley, Laloyaux, Lopez, Lupu, Radnoti,
  de~Rosnay, Rozum, Vamborg, Villaume, and Thépaut}]{era5}
Hersbach H, Bell B, Berrisford P, et~al (2020) {The ERA5 global reanalysis}.
  Quarterly Journal of the Royal Meteorological Society 146(730):1999--2049.
  \doi{10.1002/qj.3803}

\bibitem[{IPCC(2021)}]{ipcc_ar6_spm}
IPCC (2021) Summary for {P}olicymakers. In: Climate Change 2021 – The
  Physical Science Basis: Working Group I Contribution to the Sixth Assessment
  Report of the Intergovernmental Panel on Climate Change. Cambridge University
  Press, p 3–32, \doi{10.1017/9781009157896.001}

\bibitem[{Jankov et~al(2019)Jankov, Beck, Wolff, Harrold, Olson, Smirnova,
  Alexander, and Berner}]{jankov2019}
Jankov I, Beck J, Wolff J, et~al (2019) {Stochastically Perturbed
  Parameterizations in an HRRR-Based Ensemble}. Monthly Weather Review
  147(1):153 -- 173. \doi{10.1175/MWR-D-18-0092.1}

\bibitem[{Kain et~al(2017)Kain, Willington, Clark, Weiss, Weeks, Jirak,
  Coniglio, Roberts, Karstens, Wilkinson, Knopfmeier, Lean, Ellam, Hanley,
  North, and Suri}]{hwt2017}
Kain JS, Willington S, Clark AJ, et~al (2017) Collaborative efforts between the
  {United States} and {United Kingdom} to advance prediction of high-impact
  weather. Bull Amer Meteor Soc 98:937--948. \doi{10.1175/BAMS-D-15-00199.1}

\bibitem[{Keat et~al(2019)Keat, Stein, Phaduli, Landman, Becker, Bopape,
  Hanley, Lean, and Webster}]{keat2019}
Keat WJ, Stein THM, Phaduli E, et~al (2019) Convective initiation and storm
  life cycles in convection‐permitting simulations of the {Met Office Unified
  Model} over {South Africa}. Quart J Roy Meteor Soc 145(721):1323--1336.
  \doi{10.1002/qj.3487}

\bibitem[{Kendon et~al(2017)Kendon, Ban, Roberts, Fowler, Roberts, Chan, Evans,
  Fosser, and Wilkinson}]{kendon2017}
Kendon EJ, Ban N, Roberts NM, et~al (2017) Do convection-permitting regional
  climate models improve projections of future precipitation change? Bull Amer
  Meteor Soc 98:79--93. \doi{10.1175/BAMS-D-15-0004.1}

\bibitem[{Krueger(2000)}]{krueger}
Krueger SK (2000) Cloud system modeling. In: {Genereal Circulation Model
  Development}. Academic Press, pp 605--640, {ISBN}: 0-12-578010-9

\bibitem[{Lagerquist et~al(2021)Lagerquist, Turner, Ebert-Uphoff, Stewart, and
  Hagerty}]{Lagerquist2021}
Lagerquist R, Turner D, Ebert-Uphoff I, et~al (2021) {Using Deep Learning to
  Emulate and Accelerate a Radiative Transfer Model}. Journal of Atmospheric
  and Oceanic Technology 38(10):1673 -- 1696. \doi{10.1175/JTECH-D-21-0007.1}

\bibitem[{Lean et~al(2008)Lean, Clark, Dixon, Roberts, Fitch, Forbes, and
  Halliwell}]{lean_high_res_um}
Lean HW, Clark PA, Dixon M, et~al (2008) Characteristics of high-resolution
  versions of the {Met Office Unified Model} for forecasting convection over
  the {United Kingdom}. Mon Wea Rev 136:3408--3424. \doi{10.1175/2008MWR2332.1}

\bibitem[{Molteni(2003)}]{molteni2003}
Molteni F (2003) {Atmospheric simulations using a GCM with simplified physical
  parametrizations. I: Model climatology and variability in multi-decadal
  experiments}. Climate Dynamics 20(2-3):175--191.
  \doi{10.1007/s00382-002-0268-2}

\bibitem[{Orr et~al(2013)Orr, Phillips, Webster, Elvidge, Weeks, Hosking, and
  Turner}]{orr2013}
Orr A, Phillips T, Webster S, et~al (2013) {Met Office Unified Model}
  high‐resolution simulations of a strong wind event in {A}ntarctica. Quart J
  Roy Meteor Soc 140(684):2287--2297. \doi{10.1002/qj.2296}

\bibitem[{Pincus et~al(2003)Pincus, Barker, and Morcrette}]{pbm03}
Pincus R, Barker HW, Morcrette JJ (2003) A fast, flexible, approximate
  technique for computing radiative transfer in inhomogeneous cloud fields. J
  Geophys Res 108(D13). \doi{10.1029/2002JD003322}

\bibitem[{Qin et~al(2022)Qin, Zelinka, and Klein}]{Qin2022}
Qin Y, Zelinka MD, Klein SA (2022) {On the Correspondence Between
  Atmosphere-Only and Coupled Simulations for Radiative Feedbacks and Forcing
  From CO2}. Journal of Geophysical Research: Atmospheres 127(3):e2021JD035460.
  \doi{10.1029/2021JD035460}

\bibitem[{Salmanidou et~al(2021)Salmanidou, Beck, Pazak, and
  Guillas}]{salmanidou2021probabilistic}
Salmanidou DM, Beck J, Pazak P, et~al (2021) Probabilistic, high-resolution
  tsunami predictions in northern {C}ascadia by exploiting sequential design
  for efficient emulation. Natural Hazards and Earth System Sciences
  21(12):3789--3807. \doi{10.5194/nhess-21-3789-2021}

\bibitem[{Shamekh et~al(2023)Shamekh, Lamb, Huang, and Gentine}]{Shamekh2023}
Shamekh S, Lamb KD, Huang Y, et~al (2023) {Implicit learning of convective
  organization explains precipitation stochasticity}. Proceedings of the
  National Academy of Sciences of the United States of America 120(20):1--11.
  \doi{10.1073/pnas.2216158120}

\bibitem[{Slingo et~al(2022)Slingo, Bates, Bauer, Belcher, Palmer, Stephens,
  Stevens, Stocker, and Teutsch}]{slingo22}
Slingo J, Bates P, Bauer P, et~al (2022) Ambitious partnership needed for
  reliable climate prediction. Nature Climate Change 12:499--503.
  \doi{10.1038/s41558-022-01384-8}

\bibitem[{Smith(1990)}]{smith90}
Smith RNB (1990) A scheme for predicting layer clouds and their water content
  in a general circulation model. Quart J Roy Meteor Soc 116:435--460.
  \doi{10.1002/qj.49711649210}

\bibitem[{Song et~al(2019)Song, Lim, and Joo}]{song2018}
Song HJ, Lim B, Joo S (2019) Evaluation of rainfall forecasts with heavy rain
  types in the high-resolution {Unified Model} over {South Korea}. Wea
  Forecasting 34:1277--1293. \doi{10.1175/WAF-D-18-0140.1}

\bibitem[{Stein et~al(2015)Stein, Parker, Hogan, Birch, Holloway, Lister,
  Marsham, and Woolnough}]{stein2015}
Stein THM, Parker DJ, Hogan RJ, et~al (2015) The representation of the {W}est
  {A}frican monsoon vertical cloud structure in the {M}et {O}ffice {U}nified
  {M}odel: an evaluation with {CloudSat}. Quart J Roy Meteor Soc
  141(693):3312--3324. \doi{10.1002/qj.2614}

\bibitem[{Stensrud(2007)}]{stensrud}
Stensrud DJ (2007) Parameterization Schemes: Keys to Understanding Numerical
  Weather Prediction Models. Cambridge University Press,
  \doi{10.1017/CBO9780511812590}

\bibitem[{Tian and Dong(2020)}]{tian20}
Tian B, Dong X (2020) The {Double-ITCZ} bias in {CMIP3}, {CMIP5} and {CMIP6}
  models based on annual mean precipitation. Geophysical Research Letters 47.
  \doi{10.1029/2020GL087232}

\bibitem[{Tompkins(2005)}]{tompkins2005}
Tompkins AM (2005) The parametrization of cloud cover. {European Centre for
  Medium-Range Weather Forecasts, Numerical Weather Prediction Course:
  Parameterization of diabatic processes}

\bibitem[{Ukkonen et~al(2020)Ukkonen, Pincus, Hogan, and Nielsen}]{Ukkonen2020}
Ukkonen P, Pincus R, Hogan RJ, et~al (2020) {Accelerating radiation
  computations for dynamical models with targeted machine learning and code
  optimization.} Journal of Advances in Modeling Earth Systems (12).
  \doi{10.1029/2020MS002226}

\bibitem[{Vallis(2017)}]{Vallis_2017}
Vallis GK (2017) Atmospheric and Oceanic Fluid Dynamics: Fundamentals and
  Large-Scale Circulation, 2nd edn. Cambridge University Press

\bibitem[{Van~Weverberg et~al(2016)Van~Weverberg, Boutle, Morcrette, and
  Newsom}]{tke_rhcrit}
Van~Weverberg K, Boutle IA, Morcrette CJ, et~al (2016) {Towards Retrieving
  Critical Relative Humidity from Ground-Based Remote-Sensing Observations}.
  Quart J Roy Meteor Soc 142:2867--2881. \doi{10.1002/qj.2874}

\bibitem[{Walters et~al(2017)Walters, Brooks, Boutle, Melvin, Stratton, Vosper,
  Wells, Williams, Wood, Allen, Bushell, Copsey, Earnshaw, Edwards, Gross,
  Hardiman, Harris, Heming, Klingaman, Levine, Manners, Martin, Milton,
  Mittermaier, Morcrette, Riddick, Roberts, Sanchez, Selwood, Stirling, Smith,
  Suri, Tennant, Vidale, Wilkinson, Willett, Woolnough, and Xavier}]{ga6p0}
Walters D, Brooks M, Boutle I, et~al (2017) {The Met Office Unified Model
  Global Atmosphere 6.0/6.1 and JULES Global Land 6.0/6.1 configurations}.
  Geosci Model Dev 10:1487--1520. \doi{10.5194/gmd-10-1487-2017}

\bibitem[{Walters et~al(2011)Walters, Best, Bushell, Copsey, Edwards, Falloon,
  Harris, Lock, Manners, Morcrette, Roberts, Stratton, Webster, Wilkinson, R.,
  Boutle, Earnshaw, Hill, MacLachlan, Martin, Moufouma-Okia, Palmer, Petch,
  Rooney, Scaife, and Williams}]{ga3p0}
Walters DN, Best MJ, Bushell AC, et~al (2011) {The Met Office Unified Model
  Global Atmosphere 3.0/3.1 and JULES Global Land 3.0/3.1 configurations}.
  Geosci Model Dev 4:919--941. \doi{10.5194/gmd-4-919-2011}

\bibitem[{Walters et~al(2019)Walters, Baran, Boutle, Brooks, Earnshaw, Edwards,
  Furtado, Hill, Lock, Manners, Morcrette, Mulcahy, Sanchez, Smith, Stratton,
  Tennant, Tomassini, Weverberg, Vosper, Willett, Browse, Bushell, Dalvi,
  Essery, Gedney, Hardiman, Johnson, Johnson, Jones, Mann, Milton, Rumbold,
  Sellar, Ujiie, Whitall, Williams, and Zerroukat}]{ga7p0}
Walters DN, Baran A, Boutle I, et~al (2019) {The Met Office Unified Model
  Global Atmosphere 7.0 and JULES Global Land 7.0 configurations}. Geosci Model
  Dev 12:1909--1963. \doi{10.5194/gmd-12-1909-2019}

\bibitem[{Watson-Parris(2021)}]{watson2021machine}
Watson-Parris D (2021) Machine learning for weather and climate are worlds
  apart. Philosophical Transactions of the Royal Society A 379(2194):20200098.
  \doi{10.1098/rsta.2020.0098}

\bibitem[{Webster et~al(2008)Webster, Uddstrom, Oliver, and
  Vosper}]{webster_nesting_suite}
Webster S, Uddstrom M, Oliver H, et~al (2008) A high resolution modelling case
  study of a severe weather event over {N}ew {Z}ealand. Atmos Sci Let
  9(3):119--128. \doi{10.1002/asl.172}

\end{thebibliography}

\end{document}